\documentstyle[aps,prb, preprint]{revtex}

\begin{document}
\draft
\title{Infrared and Raman studies of the Verwey transition in magnetite.}
\author{L.V. Gasparov and D.B. Tanner}
\address{University of Florida, Department of Physics, POBox 118440\\
Gainesville Fl 32611-8440, USA}
\author{D. B. Romero}
\address{NIST-Optical Technology Division, Gaithersburg, MD 20899-8441
and Physics\\ Department, University of Maryland, College Park, MD 20742}
\author{H.D. Drew}
\address{ Physics Department, University of Maryland, College Park, MD 20742}
\author{H. Berger and G. Margaritondo}
\address{Physics Department-IPA, \'{E}cole Polytechnique F\'{e}d\'{e}ral de
Lausanne,%
\\
CH-1015 Lausanne, Switzerland}
\author{L. Forr\'{o}}
\address{Physics Department-IGA, \'{E}cole Polytechnique F\'{e}d\'{e}ral de
Lausanne,%
\\
CH-1015 Lausanne, Switzerland}

\date{\today}
\maketitle

\begin{abstract}

We present infrared and Raman measurements of magnetite
(Fe$_{3}$O$_{4}$). This material is known to undergo a metal-insulator
and a structural transition (Verwey transition) at $T_{\text{V}}=120$~K.
At temperatures below $T_{\text{V}}$, we observe a strong gap-like
suppression of the optical conductivity below 1000 cm$^{-1}$. The
structural aspect of the Verwey transition demonstrates itself by the
appearance of additional infrared- and Raman-active phonons. The
frequencies of the infrared-active phonons show no significant
singularities at the transition whereas their linewidths increase. The
frequency and linewidth of the Raman-active phonon at 670 cm$^{-1}$
changes abruptly at the transition. For $T<T_{\text{V}}$, we observe
fine structures in the infrared and Raman spectra which may indicate
strong anharmonicity of the system below the transition. Our estimate of
the effective mass of the carriers above the transition to be $m^* 
\approx 100 m$, where $m$ is the electron mass.
Our measurements favor a polaronic mechanism of conductivity and
underline the importance of the electron-phonon interaction in the
mechanism of the Verwey transition. 
\end{abstract}


\section{Introduction}

\subsection{The Verwey transition}

Charge ordering in strongly-correlated electron systems has recently
attracted a lot of attention on account of its possible influence on
charge transport and dynamics. In the high-temperature superconductors,
there is evidence for the formation of charge and spin stripes near the
commensurate doping $x=1/8$ in La$_{2-x}$Sr$_{x}$CuO$_{4}$ in which
superconductivity is suppressed.\cite{Axe89} The suppression was
attributed to the pinning by lattice distortions of the dynamic charge
and spin stripes formed in this phase.\cite{Tranquada97}
Very recently Raman scattering in NaV$_2$O$_5$ compound 
revealed considerable charge ordering there at T$<$35 K.\cite{Zgenya99} 
In the
manganates, charge ordering is observed at certain commensurate 
doping.\cite{Mori98} The possibility of the coexistence of ferromagnetic and
charge-order correlations may be relevant in understanding the
phenomenon of colossal magnetoresistance in manganates. The earliest
compound known\cite{Ver39} to manifest charge ordering is magnetite
(Fe$_{3}$O$_{4}$).

Fe$_{3}$O$_{4}$ is a ferrimagnet below 850~K. At room temperature
Fe$_{3}$O$_{4}$ has an inverse spinel cubic structure with the
$O_{h}^{7}$ ($Fd3m$) space group.\cite{Brag15} The unit cell consists
of 8 formula units. There are two types of Fe positions in this
structure. The so-called A positions are characterized by the
tetrahedral oxygen surrounding the Fe ions; the B positions have 
octahedral oxygen coordination. The A positions are occupied only by
Fe$^{+3}$ ions, whereas the B positions are occupied with equal
probability by Fe$^{+2}$ and Fe$^{+3}$.

Magnetite undergoes a first-order transition (the Verwey transition)
at $T_{\text V} = 120$ K, with changes of
crystal structure, latent heat, and a two-order of magnitude
decrease of the DC conductivity. This transition was discovered by
Verwey\cite{Ver39} in 1939; in his model, the low temperature phase
has the Fe$^{+2}$ and Fe$^{+3}$ ions at the B-sites ordered, forming
layers of ions of each charge state. Above $T_{\text V}$ there is no
long-range order, but short-range order persists. According to neutron
measurements,\cite{Izum82} an orthorhombic space groups (either a
D$_{2h}^{11},(Pmca)$ or C$_{2v}^{2},(Pmc2_{1})$) may describe the low
temperature structure.

The Verwey transition has been intensively investigated since its
discovery, bur remains unexplained.
Basically, there are several ideas which partly explain experimental
results although none is able to give a complete explanation of the
transition. This incomplete theory is probably not surprising in a
system where the electron-phonon interaction, the electron-electron
interaction, and electronic bandwidth are of nearly the same importance. We
shall mention some basic ideas which could lead to an understanding of
the transition.

As already mentioned, Verwey and Haayman\cite{Verv41} originally
described this transition as an order-disorder transition. The ordered
state is insulating, whereas the disordered state is metallic
due to the motion of the extra electron on the B-positions. Later, 
Anderson\cite{Ander56} argued that short range order (SRO) is crucial
for understanding the transition. He demonstrated that certain SRO
configurations are energetically favorable (Anderson's condition). In
this picture, long range order is destroyed above the transition
although short-range order, satisfying Anderson's conditions, persists
above the transition. The presence of SRO above the transition was
confirmed by neutron measurements. However, there is still a question
why an abrupt increase in conductivity is observed in spite of the
persistence of SRO above the transition.\cite{Mott80}

Cullen and Callen\cite{Cul70} presented a picture of the transition 
in which only
electron-electron correlations are relevant. It is assumed, that
an Fe$^{+2}$ ion can be represented as an Fe$^{+3}$ ion plus an extra
electron. In this case, the physics of Fe$_{3}$O$_{4}$ is just the
physics of the interaction of this extra electron with a crystal of 
Fe$^{+3}$ ions.
The interaction 
can be described by the Coulomb nearest neighbor repulsion and
hopping integral for the extra electron. The competition between these
two values determines whether an insulator or metal is obtained.

In contrast Mott,\cite{Mott80,Aust69} Chakraverty,\cite{Chak74} and
Yamada\cite{Yam80} emphasized the importance of the electron-phonon
interaction. In the Mott\cite{Mott80} picture of the Verwey transition,
the charge carriers are either polarons or bipolarons. Above the
transition, the system can be described as a Wigner glass, a system
where carriers are localized by the disorder-induced random-field
potentials. When the temperature is raised, some of the polarons
dissociate leading to a hopping type of conductivity.

One should note however that an activated conductivity could also be
accounted for by the opening of a charge gap. Recent photoemission
studies\cite{Par97} of magnetite indicate a charge gap of 150 meV below
$T_{\text V}$.
A recent optical measurement on magnetite\cite{Par98}
found that the main temperature-dependent changes of optical
conductivity occur below 1 eV, connected with a gap-like
temperature-dependant increase of optical conductivity at 140~meV, followed
by a broad polaronic peak at 0.6 eV. The gap-like behavior was
attributed to the opening of a charge gap. The authors of
Ref.~\onlinecite{Par98} suggested that the charge gap, determined as
the onset of the polaronic band in the spectra below the transition, is
consistent with the gap value found in photoemission studies.
In contrast to the statement of Ref.~\onlinecite{Par98}, we must note
that photoemission measures the transitions to the Fermi level, whereas
infrared spectroscopy measures an optically-allowed transition from one
electronic band to another across the Fermi level. This means that the
optical gap of 140~meV would correspond to smaller gap, perhaps as small
as 70~meV, measured in photoemission. Moreover, according to
Ref.~\onlinecite{Par97}, the gap does not close above transition but
rather decreases by only 50~meV, which is not consistent with the
analysis of the optical conductivity of Ref.~\onlinecite{Par98}.




In this paper we present the results of systematic temperature-dependant
infrared and Raman measurements of magnetite and discuss both lattice
and charge dynamics, as well as a possible mechanism of
Verwey transition.

\subsection{Optical phonons in magnetite} 

Above the Verwey transition temperature, $T>T_{\text{V}}=120$ K,
magnetite has a cubic-inverse-spinel structure belonging to the space
group O$_{h}^{7}(Fd3m)$. The full unit cell contains 56 atoms but the
smallest Bravais cell contains only 14 atoms. As a result, one should
expect 42 vibrational modes. Group theory predicts the following modes
for this phase.
\[ A_{1g}+E_{g}+T_{1g}+3T_{2g}+2A_{2u}+2E_{u}+5T_{1u}+2T_{2u}. \]
The $T_{1g}$, $A_{2u}$, $E_{u}$, and $T_{2u}$ are silent. Thus, there
are five Raman-active modes ($A_{1g}+E_{g}+3T_{2g}$) and five
infrared-active modes ($5T_{1u}$).

When the system undergoes the Verwey transition, the symmetry of the
crystal lowers to orthorhombic. The unit cell doubles along the $z$
direction and the diagonals of the former cubic unit cell become the
faces of the low-temperature unit cell. As mentioned in 
Ref.~\onlinecite{Izum82}, it has not been possible to get a complete
refinement from the neutron scattering data. Nevertheless,
we can analyze the proposed space groups and compare the results of the
group theory analysis with our data. Let us start with $Pmca$ which is a
nonstandard setting of the $D_{2h}^{11},Pbcm$ (No.~57). The unit cell
contains 56 atoms, and in this case we cannot reduce the unit cell as
was possible for the room temperature crystal structure. The expected
number of modes is 168. Group theory predicts the following modes for
this phase 
\[23A_{g}+18A_{u}+24B_{1g}+19B_{1u}+16B_{2g}+27B_{2u}+15B_{3g}+26B_{3u} \]
The $A_{u}$ modes are silent. Thus, there are 78 Raman-active modes
($23A_{g}+24B_{1g}+16B_{2g}+15B_{3g}$) and 72 infrared-active modes
($19B_{1u}+27B_{2u}+26B_{3u}$). It is clear from this analysis that one
should expect a dramatic increase in the number of modes in the low
temperature phase. Some of the new modes are the former
Brillouin-zone-boundary modes which become visible because of the unit
cell doubling; others are due to the lifting of degeneracy of former $T$
and $E$-type modes. Some modes persist over the transition. For
instance, the $A_{1g}$ mode of the cubic phase becomes an $A_{g}$ mode
of the orthorhombic phase, having the same Raman tensor.

The second proposed symmetry is $C_{2v}^{2},Pmc2_{1}$, (No.~26). This
group does not contain an inversion center, leading to very important
consequences for the Raman and infrared spectra. Namely, if magnetite
has such a structure below $T_{\text{V}}$, both infrared-active and 
Raman-active type modes 
should be detected by both spectroscopes. Experimentally, this is not
the case and we can rule this symmetry out. 

The symmetry of the Raman
modes can be figured out by their polarization selection rules. The
$A_{g}$ and $E_{g}$-type modes have diagonal Raman tensor so that they
are seen only when the polarizations of the incident and scattered light
are parallel. In contrast, the $T_{2g}$, $B_{1g}$, $B_{2g}$ and $B_{3g}$
modes have off-diagonal components so that they are observed when the
polarizations of the incident and scattered light are perpendicular.

\section{Experimental}

The magnetite single crystals used in this work were grown by a chemical
vapor transport technique using stoichiometric Fe$_{3}$O$_{4}$
microcrystalline powder obtained by reduction reaction of ferric oxide
(Fe$_{2}$O$_{3}$ ). This procedure yielded near-stoichiometric
crystals with typical size of 4x4x2~mm$^{3}$. X-ray diffraction
confirmed the spinel-type structure of the crystals. Transport
measurements found the abrupt increase in resistance below $T=120$K,
characteristic of the Verwey transition. Optical measurements were
carried out on the optically smooth surfaces of the as-grown single
crystals.

The reflectance was measured in the frequency range of 50--45000~cm$^{-1}$ 
(6 meV--5.5 eV)
at several temperatures between 40 and 300~K. The
sample temperature was maintained by mounting the crystal on the
cold-finger of a He-flow cryostat. To cover a wide frequency range, we used
three optical set-ups. A Bruker IFS 113V spectrometer was employed for
the far infrared and mid infrared regions (50--4500 cm$^{-1}$; 6--550 
meV), a Perkin
Elmer 16U grating spectrometer for the near infrared and visible region
(3800--24000~cm$^{-1}$; 0.47--3 eV). We observed no temperature dependence
of the
spectra above 20000 cm$^{-1}$. Therefore, in order to extend the data to
higher frequencies (20000-45000 cm$^{-1}$; 2.5--5.5 eV) 
we merged the spectrum at a
given temperature with the room temperature reflectance spectrum
obtained from a Zeiss grating spectrometer coupled with microscope, and
finally in the frequency range above 45000 cm$^{-1}$ we merged our data
with the data of Ref.~\onlinecite{Par98}.

In order to analyze the optical properties of the sample we performed
Kramers-Kronig transformation of reflectance data. For the temperatures
above the transition, we use Drude-Lorentz model fit in the low
frequency part of the spectrum. Then we calculated the reflectance and
used the result for the low-frequency extrapolation. Below the
transition, we assume constant DC conductivity at zero frequency as a
low frequency approximation. For the high frequency approximation, we
used a power law ($\omega ^{-4}$) extrapolation.

The polarized Raman scattering experiment was conducted a wide range of
temperatures ($T=5$--$300$ K) using a triple spectrometer equipped
with a liquid-nitrogen cooled CCD detector. The 514.5 nm line of an
Ar$^{+}$ laser was used as excitation with nominally 25 mW incident on
the sample. The spectra were taken in the back-scattering geometry with
the scattered light polarized vertically while the incident light is
polarized to select a particular component of the Raman scattering
tensor.

\section{RESULTS AND DISCUSSION}

\subsection{Infrared and Raman-active Phonons}

We now proceed to analyze the infrared and Raman data keeping in mind
all of the above-predicted changes. The infrared reflectance and optical
conductivity are shown in Fig.~1 and 2, while the Raman-scattering
spectra of Fe$_{3}$O$_{4}$ are in Fig.~3. Oxygen phonon modes usually
have their frequency above 200 cm$^{-1}$, whereas vibrations of the
heavier iron atoms should be at much smaller frequency. In Fig.~1 and 2,
two $T_{1u}$ oxygen modes are found near 350 cm$^{-1}$ and 560
cm$^{-1}$. These peaks are similar to the infrared-active phonons
reported earlier by different groups.\cite{Shleg79,Shleg80,Deg87,Degi87} 
Four Raman-active phonons are
observed at room temperature in Fig.~3. The strongest peak at
$\approx$670~cm$^{-1}$ is either the $A_{1g}$ or the $E_{g}$ mode since
it is only present in the parallel polarization geometry (XX). Since we
expect the $A_{1g}$ mode to have the highest frequency as it involves
the stretching vibrations of the oxygen atoms along the Fe(A)-O bonds,
the 670~cm$^{-1}$ phonon is assigned to the $A_{1g}$ mode. In Fig. 3,
Raman spectrum in which the incident and scattered light polarizations
are perpendicular contains the predicted three $T_{2g}$ modes at 193
cm$^{-1}$, 308~cm$^{-1}$, and 540~cm$^{-1}$. Verble\cite{Verble74} made
a similar assignment of the Raman-active phonons in Fe$_{3}$O$_{4}$
except for the 193~cm$^{-1}$ which he did not observe. Instead, he
reported that the lowest frequency $T_{2g}$ mode is at 300~cm$^{-1}$ as
observed in the spectrum at $T=77$ K. However, this feature seen by
Verble at $T=77$ K ($<T_{\text{V}}$) is related to the structural
symmetry-breaking that occurs below $T_{\text{V}}$ as discussed below.

The Verwey transition manifests itself in both infrared and Raman
spectra by remarkable changes in the phonon spectrum as the temperature
is lowered through $T_{\text{V}}$. In the Raman spectrum, one can see at
least 17 new modes below $T_{\text{V}}$. In addition to the new phonon
modes there is a broad (about 150~cm$^{-1}$, full width on half maximum)
peak with a maximum at 350~cm$^{-1}$. Unfortunately, with so many
additional phonon modes it is difficult to judge whether this broad peak
is a superposition of phonon modes or a background on which phonon modes
are sitting.

There are several additional infrared modes as well. In Fig. 1 and 2b,
in addition to the infrared active phonons near 350~cm$^{-1}$ and 
560~cm$^{-1}$, weak phonon features appear below 300~cm$^{-1}$. Similar
effects are observed in the Raman spectrum at $T=5$ K as illustrated in
Fig.~3. Moreover, the Verwey transition also leads to renormalization of
the phonon frequencies and linewidths. The infrared- and Raman-active
phonons which persist over the transition were fitted to Lorentzian
lineshapes in order to obtain their frequency and linewidths. The
results of this analysis are shown in Fig. 4. For the infrared active
phonons at 350~cm$^{-1}$ and 560~cm$^{-1}$, the phonon frequencies
increase upon decreasing the temperature but without visible anomalies
around $T_{\text{V}}$. On the other hand, their linewidths manifest an
abrupt increase near $T_{\text{V}}$. Additionally, for the 350 and 
560~cm$^{-1}$ phonons there is a reproducible fine structure in the
reflectance; see Fig.~1b. At temperatures below the transition the
lineshape of the 350 and 560~cm$^{-1}$ phonons can not be 
described by a single Lorentzian, which leads to bigger error bars for
the linewidth and frequency of these phonons below the transition. See
Fig.~4. Note that in contrast to the Chakraverty\cite{Chak74} result,
within our experimental error we do not observe any soft optical mode.

The observed phonon anomalies can be explained in the following way.
Above the transition most of the phonon modes are highly degenerate
modes. When magnetite undergoes the Verwey transition, the crystal
symmetry reduces to orthorhombic, lifting the degeneracy of the modes. 
We suggest that splitting, determined by the degree of
distortion associated with the symmetry changes, will not be
too large. According to the neutron
data\cite{Izum82} the Fe-O distances for both A and B sites are
generally within the estimated standard deviations for the prototype
structure; therefore, one expects the splitting to be small,
appearing as a broadening of the modes above the transition. This is
what has been observed in our experiment for the two infrared active
modes at 350 and 560~cm$^{-1}$.

The frequency and linewidth of the Raman-active phonon at 670~cm$^{-1}$
in Fig.~4 shows an abrupt increase in frequency with a concomitant
precipitous drop in the linewidth at $T_{\text{V}}$. The symmetry of
this mode in the cubic phase is $A_{1g}$; it corresponds to the in-phase
vibrations of the oxygen ions forming tetrahedra surrounding the Fe ions
in the A positions.\cite{Verble74} At the same time, a similar type of
displacement for the cubic-orthorhombic structural transition has been
found from neutron measurements.\cite{Izum82} Therefore, it is
plausible to assume that 670~cm$^{-1}$ phonon is directly coupled to the
structural displacements at the Verwey transition. This is the reason
why this mode is so sensitive to the transition.

\subsection{Charge Dynamics}

At room temperature, in addition to the phonon features, one notes a
gradual decline of reflectance (Fig. 1a) towards higher frequencies with
a distinctive minimum at about 10000~cm$^{-1}$ and some weak and almost
temperature independent transitions occurring in the visible and ultra
violet region.

With decreasing temperature, the low frequency reflectance (below 
1000~cm$^{-1}$) decreases. Below $T_{\text{V}}$ a minimum forms at $\approx$
750~cm$^{-1}$, followed by increased reflectance, with a temperature-
dependant hump around 4000~cm$^{-1}$. Above 20000~cm$^{-1}$, the
reflectance is practically temperature independent.

The optical conductivity derived from the reflectance data is shown in
Fig. 2. Aside from the phonon peaks, one notes a decrease of the low
frequency conductivity (below 1000~cm$^{-1}$) with decreased temperature
as well as a broad temperature-dependent band at 5000~cm$^{-1}$. The
intensity of this band increases with decreased temperature followed by 
slight temperature dependent increase of the conductivity towards higher
frequencies. The most interesting aspect of the data is the
temperature-dependence, which we will discuss next.

We will base our discussion on the local density approximation
calculations of Ref.~\onlinecite{Zha91}. According to these calculations
of the single-electron density of states, there are O($2p$) bands lying
about 4~eV below Fermi level which is found in the middle of the
Fe($3d$) bands. Therefore, the interesting physics relevant to the
Verwey transition is connected with these Fe($3d$) bands. Within the
Fe($3d$) bands one should pay special attention to the two
spin-polarized bands. Let us recall that there are two magnetic
sublattices corresponding to the A and B positions in the unit cell,
with the spins parallel to each other within one sublattice and
antiparallel to the spins in another sublattice. From the electronic
structure,\cite{Zha91} one expects a peak in the optical conductivity
associated with the optical transition from B site to B site. It is
clear that such a peak should be sensitive to the Verwey transition
because the B sites are the sites where ordering takes place. A good
candidate for this peak is the 5000~cm$^{-1}$ peak, which changes in
intensity with temperature. In the framework of the Mott model\cite{Mott80} 
of the Verwey transition this is the polaronic peak. The
increase in oscillator strength above 10000~cm$^{-1}$ is attributed to
the oxygen-iron optical transitions, although some optically allowed
transitions from A to B site may also contribute.

Another striking feature which should be explained is a gap-like feature
in the optical conductivity at~1000 cm$^{-1}$ at temperatures below the
Verwey transition. To analyze changes of conductivity we estimate the
oscillator strength or spectral weight in the low-energy region. The
effective number of carriers (per Fe$^{+2}$ ion) participating in
optical transitions at frequency less than $\omega$ is defined as\cite{Woot72}
\begin{displaymath}
\frac{m}{m^*} N_{eff}(\omega) =
\frac{2mV_{cell}}{\pi e^2 N_{{\rm Fe}^{+2}}}
\int_{0}^{\omega}\sigma_1(\omega^{'})d\omega^{'},
\end{displaymath}
where $m$ is the free-electron mass, $m^*$ is the effective mass of the
carriers, $V_{cell}$ is the unit-cell volume, and $N_{\text Fe^{+2}}$ is the
number of Fe$^{+2}$ ions per unit cell.

One can clearly see in Fig.5 an increase of $N_{eff}$ with temperature.
For temperatures below the transition, we found the maximum change in 
this
quantity at frequencies around 1800 cm$^{-1}$. In
order to compare the  temperature dependence of $N_{eff}$ with the
DC conductivity, we show the $N_{eff}$(1800 cm$^{-1}$), for
different temperatures together with the conductivity in the insert of
Fig.5. The sharp increase of $N_{eff}$ closely resembles the very
similar increase in conductivity at the Verwey transition. A similar 
result was shown by Park et al.\cite{Par97}

One should note that there are several ways to interpret the gap-like
feature in the conductivity. One can assign such a behavior either to
an activated mobility of the form $\mu_0\exp(-W_H/k_B$T), where $W_H$
is the polaron hopping energy, or to the opening of a charge gap 
$\Delta$. In later case the carrier concentration will behave like
$n_0\exp(-\Delta/k_B$T), leading to activated conductivity as well. It
is clear that both effects may alter the conductivity. However, the
agreement of $N_{eff}$ and the conductivity suggests activation of the
free carrier density.

Using the definition of $N_{eff}$ we can try to estimate effective mass
of the carriers. Let us assume that $N_{eff}$(1800 cm$^{-1}$) is
purely due to the free carrier optical conductivity. In this case, using
the definition of $N_{eff}$, we can expect $N_{eff}$(1800 cm$^{-1}$,
300K)$\approx m/m^*$, which gives us $m^*\approx$100$m$. Now let us
compare this number to the effective mass from Mott equation:\cite{Mott80}
\begin{displaymath}
m^* \approx 5m\exp(W_H/\frac{1}{2}\hbar\omega),
\end{displaymath}
where $\hbar\omega$ is the zero-point phonon energy.
 
We can get the value of $W_H$ from the polaron peak position. Namely,
the peak position ($\approx$ 5000 cm$^{-1}$) gives us $2 W_B$, where 
$W_B$ is the polaron binding energy. In the simple model of an electron
jumping between two molecules\cite{Aust69} the hopping energy can be
estimated as $W_H\approx1/2W_B$, yielding $W_H=1250$~cm$^{-1}$. The
zero-point phonon energy can be estimated to be the frequency of the
highest-energy optical phonon, 670 cm$^{-1}$. These estimates give
$m^*\approx 200 m$, the same order of magnitude that we estimated from
$N_{eff}$. Note that the Mott formula contains an exponent which is very
sensitive to both value of $W_H$ and $\frac{1}{2}\hbar\omega$, therefore
the fact that we have got the right order of magnitude for the effective
mass is consistent with the idea of a polaron mechanism of conductivity
in magnetite.

\section{Conclusions}

We have presented results for infrared and Raman
measurements of magnetite. Both Raman and infrared spectroscopic
techniques observed phonon anomalies at $T_{\text{V}}$ which are attributed to
the symmetry and structural changes below $T_{\text{V}}$. We are able to 
rule out a structure lacking inversion symmetry.

Our optical data seem to fit to the following description. At the 
temperatures below $T_{\text{V}}$ carriers are localized polarons and we have a
strong polaron peak at $\approx$ 5000 cm$^{-1}$ and a gaplike decrease of
optical conductivity below 1000~cm$^{-1}$. All polarons are bound and
position of the polaron peak occurs at twice the binding energy.

At temperatures above the transition, some carriers are
delocalized, leading to removal of oscillator strength from the 
5000~cm$^{-1}$ peak, broadening, and an increase of the low-frequency
conductivity. It is difficult to separate effects of the filling of the
charge gap or the activated increase of mobility. However, no matter
what is causing it, the low frequency optical conductivity follows the
DC conductivity. The increase of the 5000~cm$^{-1}$ band oscillator
strength with decrease of the temperature is consistent with the
polaronic origin of this band, consistent with the previous observations of
Park et al.\cite{Par97} The effective mass of the carriers estimated
from the optical conductivity gives a value of $\approx 100 m$, 
close to that predicted by
Mott in the framework of the polaron model of the Verwey transition. All
these observations strongly indicate the importance of lattice dynamics
effects, favoring the polaronic picture of the Verwey transition.

\section*{Acknowledgments}

Work at Florida was supported by NSF grant DMR-9705108.
L.V.G acknowledges Dr. E. Ya. Sherman for the critical discussion of
the manuscript.
D.B.R. acknowledges Dr.Raju Datla for his interest in and support of
the Raman work at NIST. H.B. wishes to express appreciation to
Dr. J. Lammer for fruitful discussions.

\begin{figure}

\caption{ a) The reflectance of magnetite in the frequency range from 50 to
45000 cm$^{-1}$ b) The low-frequency (up to 700 cm$^{-1}$) reflectance
of magnetite at temperatures above and below the Verwey transition.}
\end{figure}

\begin{figure} \caption{The optical conductivity of magnetite
calculated by the Kramers-Kronig transformation of the reflectance data.
Fig. 2a shows the optical conductivity below 45000 cm$^{-1}$ and Fig. 2b
shows the low frequency optical conductivity (up to 700 cm$^{-1}$) 
at temperatures above and below the Verwey transition.}
\end{figure}

\begin{figure} \caption{The Raman spectrum of magnetite at temperatures
above and below the Verwey transition. Shown are spectra in the
XX-geometry and XY-geometry. The insert shows the XX-spectrum below
600cm$^{-1}$.} \end{figure}

\begin{figure} \caption{The frequency (solid circles) and linewidth (opened
circles) of the infrared and Raman modes as determined by Lorentzian
fits to the data.} \end{figure}

\begin{figure} \caption{The effective number of carriers participating
in optical transitions at frequencies below $\omega$ ($N_{eff}$($\omega$))
versus $\omega$ at temperatures above and below the Verwey transition.
The insert shows $N_{eff}$ at 1800 cm$^{-1}$ versus temperature (solid
circles), and the DC conductivity versus temperature (open circles).}
\end{figure}

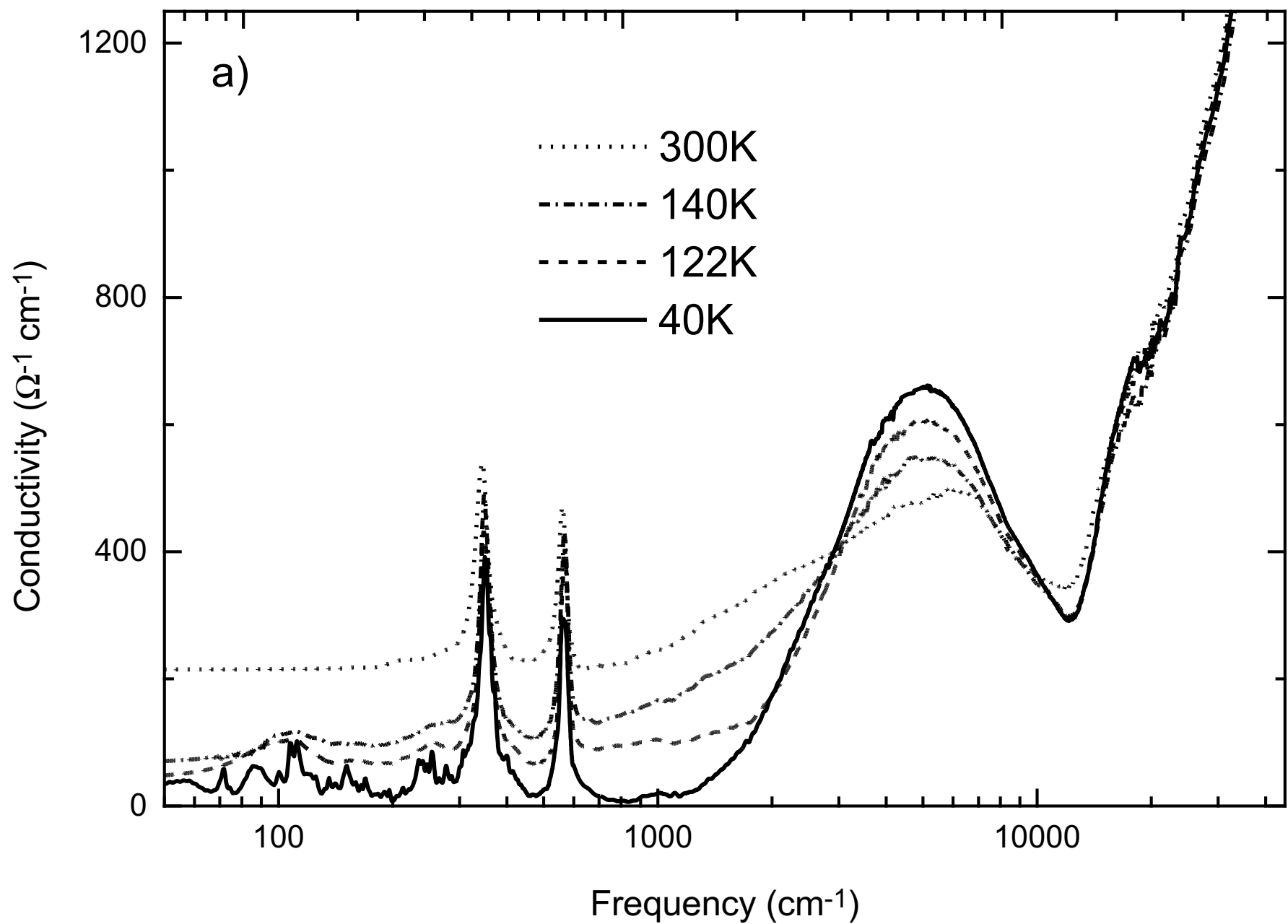
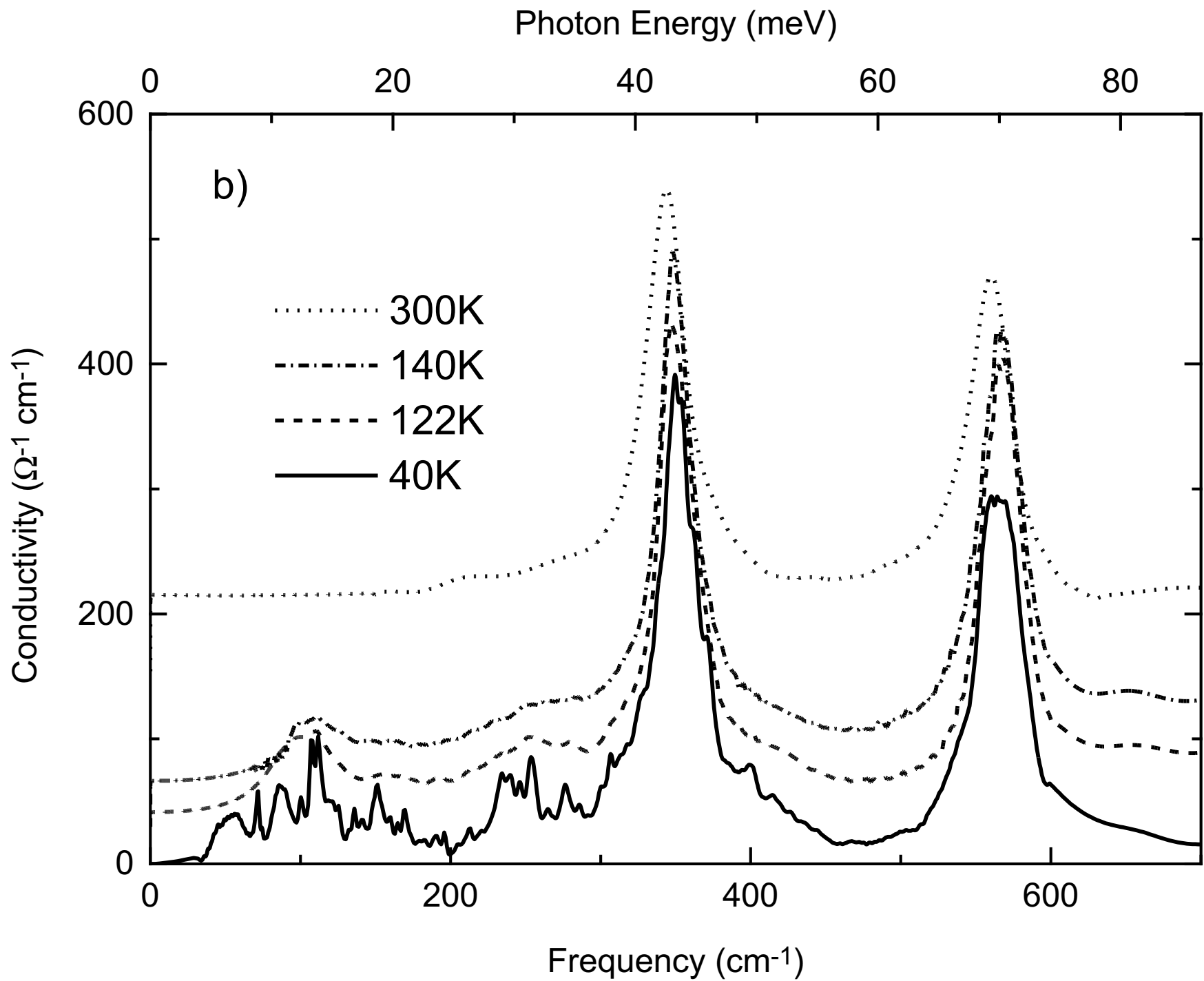
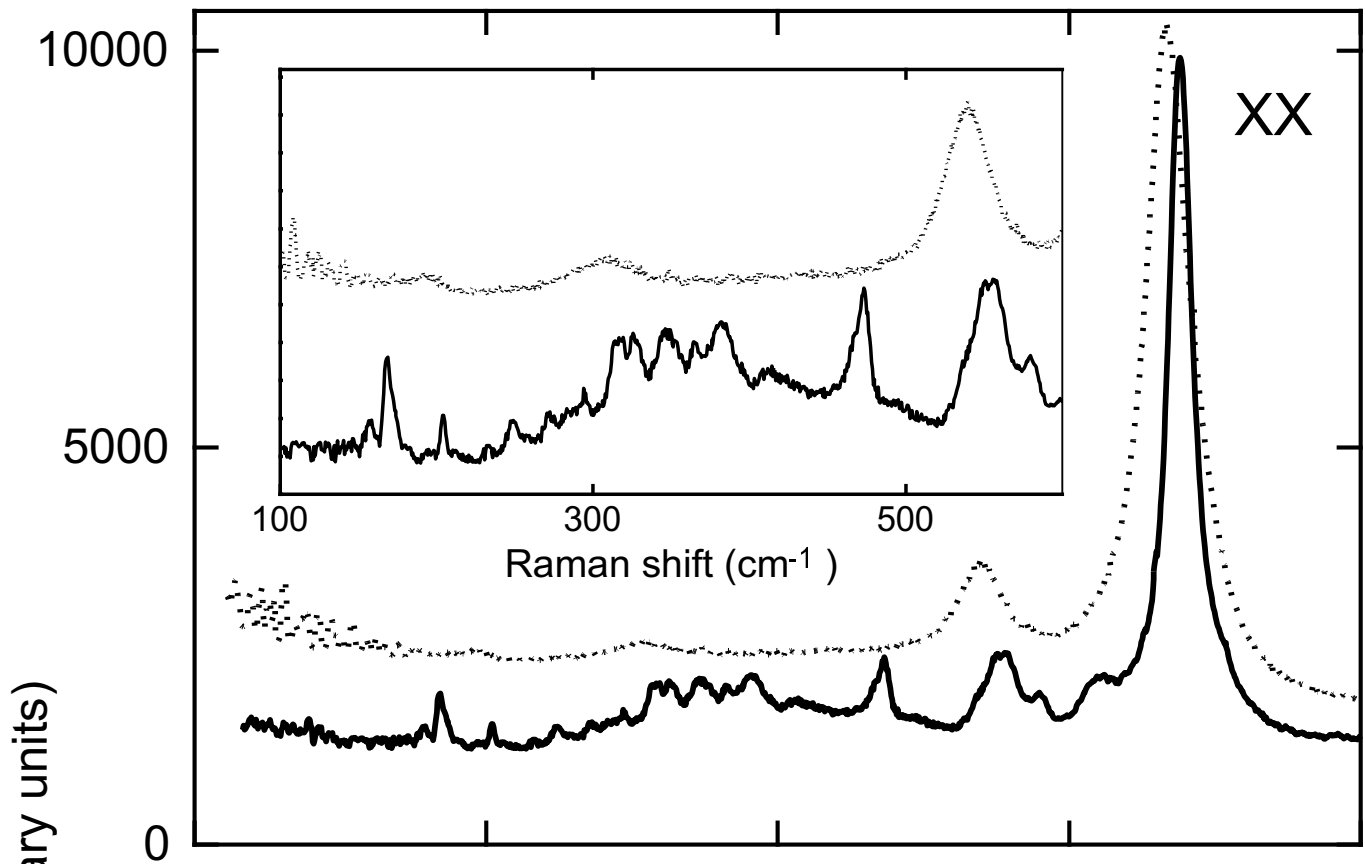
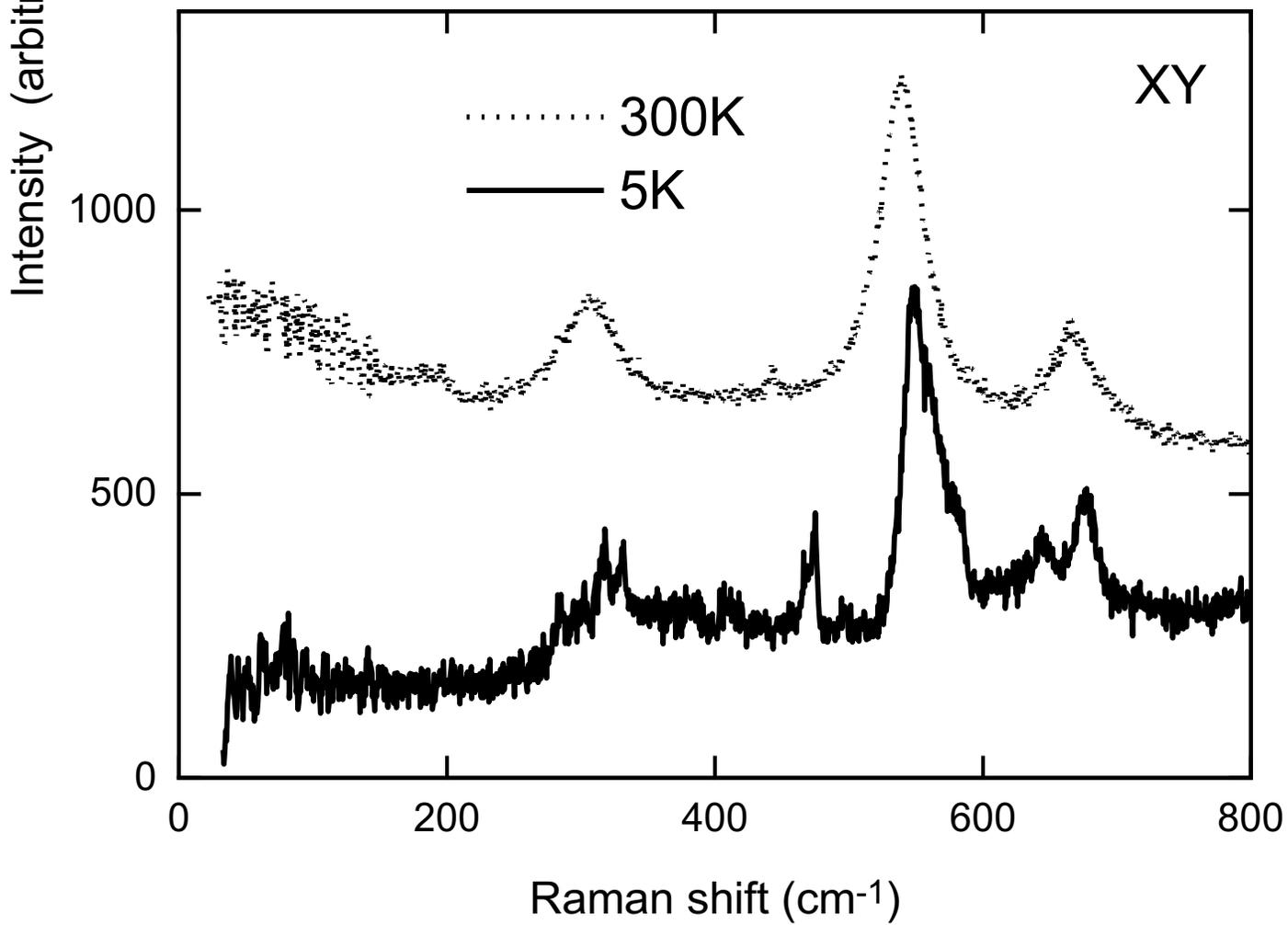
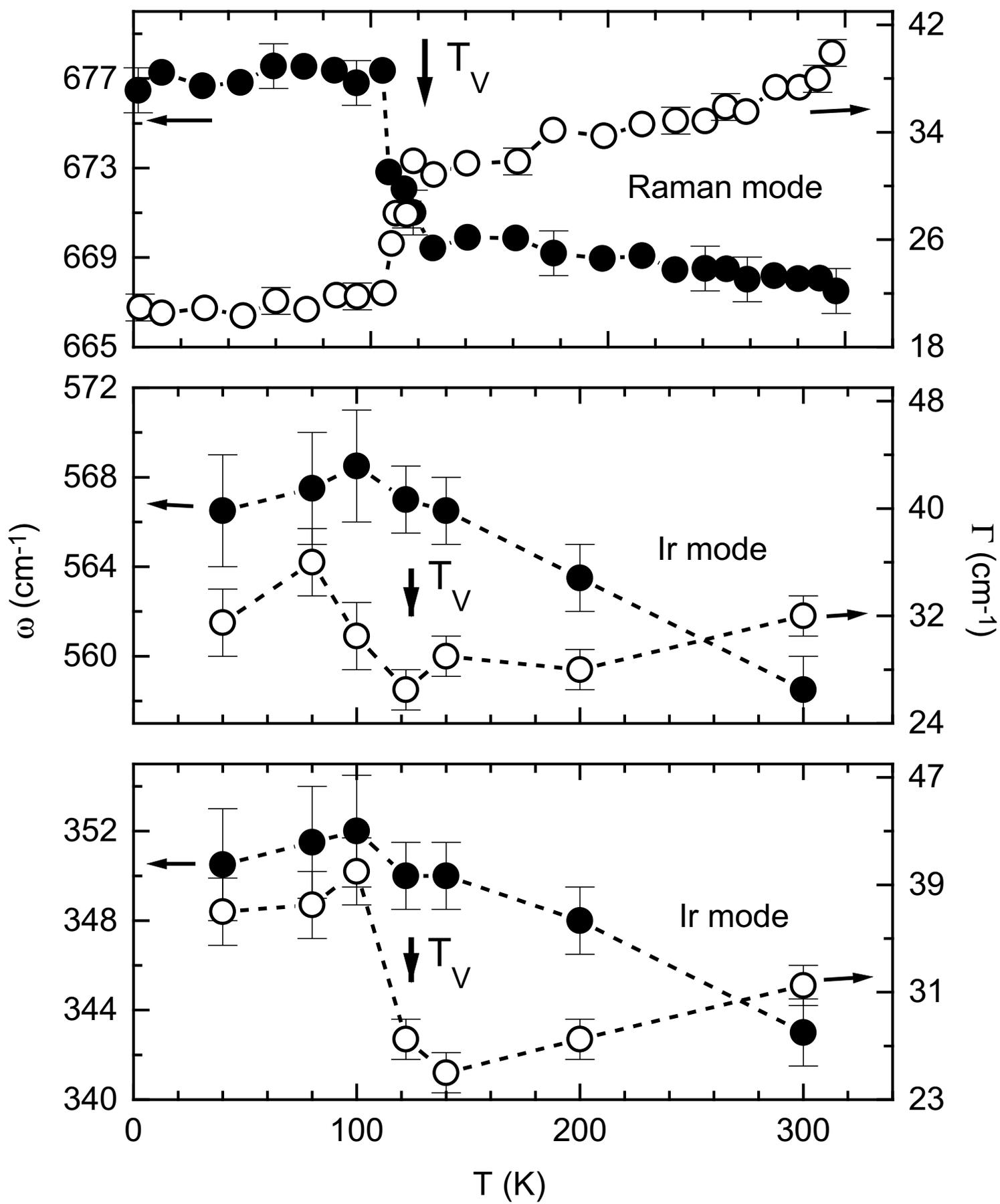
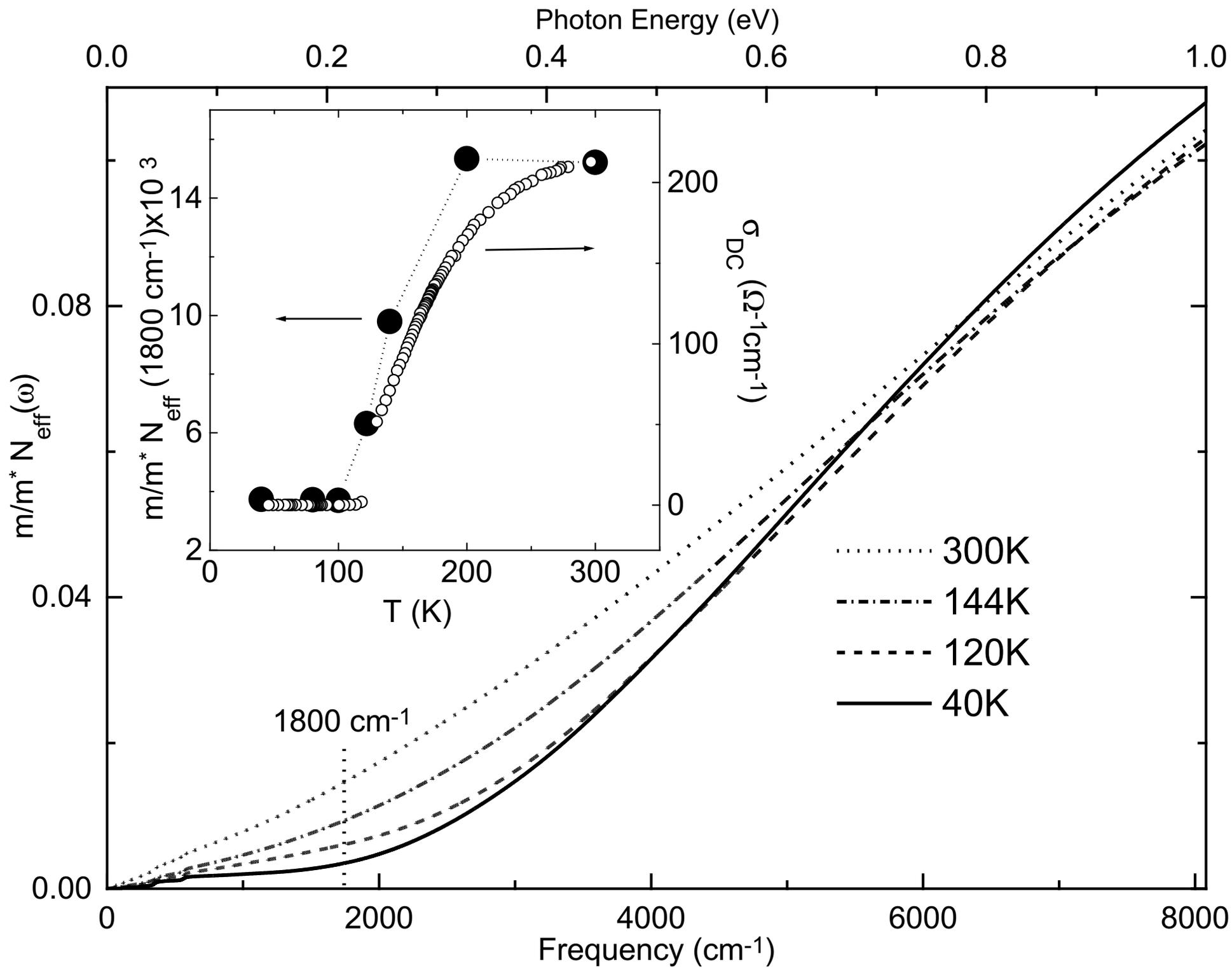
\end{document}